\documentclass[preprint2]{aastex}
\usepackage{natbib}
\begin{document}

\title{\large \rm TESTING NEW IDEAS REGARDING THE NATURE OF INTERSTELLAR EXTINCTION}
\author{\sc \small D.~G. Turner$^1$, D.~J. Majaess$^{1,2}$, D.~D. Balam$^3$}
\affil{\footnotesize $^1$ Saint Mary's University, Halifax, NS, B3H 3C3, Canada.}
\affil{\footnotesize $^2$ Mount Saint Vincent University, Halifax, NS, B3M 2J6, Canada.}
\affil{\footnotesize $^3$ Dominion Astrophysical Observatory, 5071 West Saanich Road, Victoria, British Columbia, V6A 3K7, Canada.}
\email{turner@ap.smu.ca}

\begin{abstract}
The nature of Galactic interstellar extinction is tested using reddening line parameters for several fields in conjunction with equivalent widths $W(\lambda4430)$ for the diffuse interstellar band at $4430$ \AA. The Cardelli et al.$\;$relations [29] at infrared, optical, and ultraviolet wavelengths are inconsistent with the newly-derived quadratic variation of $R_V({\rm observed})$ on reddening slope {\it X}. A minimum of $R_V=2.82\pm0.06$ exists for $X=0.83\pm0.10$, and is argued to represent true Galactic extinction described by $A(\lambda)\propto \lambda^{-1.375}$. It matches expectations for a new description of extinction in the infrared, optical, and ultraviolet by Zagury [32]. Additional consequences, reddened stars with no 2175 \AA$\;$feature and a correlation of normalized $\lambda4430$ absorption with {\it X}, are not predicted by the Cardelli et al.$\;$relation [29]. Known variations in {\it X} from 0.62 to 0.83, and corresponding variations in $R_V({\rm observed})$ from 4.0 to 2.8, presumably result from forward-scattered starlight in the ultraviolet contaminating optical light of stars affected by dust extinction. A new understanding of the true nature of interstellar extinction is important for establishing an accurate picture of the extragalactic distance scale, which in turn is related to our understanding of the nature of the Universe.
\end{abstract}

\section{{\rm \footnotesize INTRODUCTION}}
The existence of opaque dust clouds in the Milky Way blocking the light of background stars was revealed dramatically when observational astronomers exposed photographic plates to starlight originating from the direction of the Galaxy's disk in the late nineteenth century [1]. The presence of dust as a more pervasive component of the interstellar medium was recognized by astronomers doing star counts to establish the dimensions of the Galaxy [2,3,4] in the second decade of the twentieth century. General confirmation of the interstellar component of Galactic dust, however, is generally attributed to Trumpler's 1930 photometric and spectroscopic studies of the Galaxy's bright open star clusters [5]. 

\begin{figure*}[!t]
\center
\includegraphics[width=9cm]{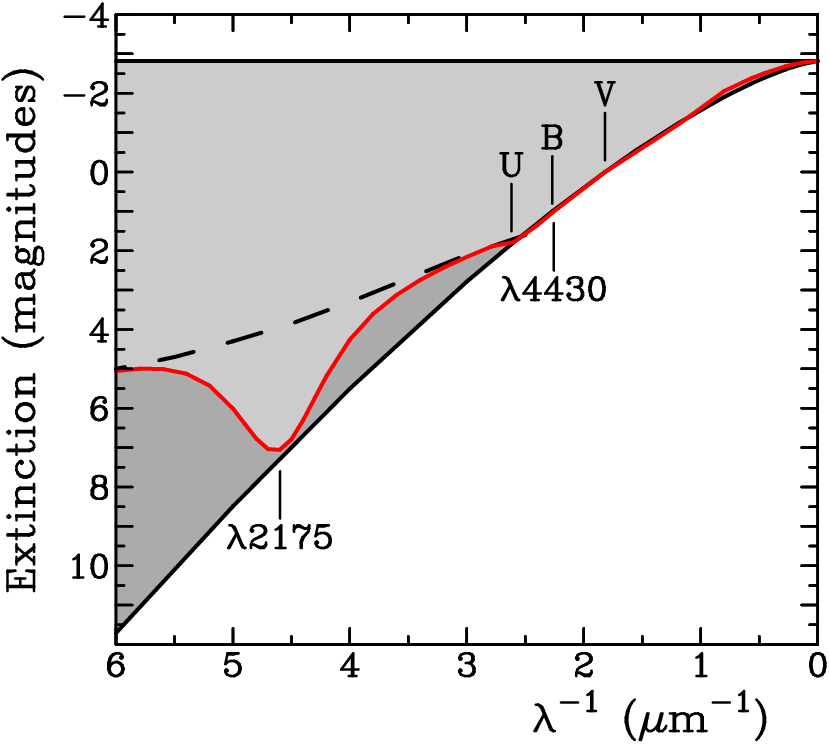}
\caption{\small{A schematic colour difference diagram illustrating a simple $\lambda^{-1.375}$ extinction law, normalized to $A_V=0$ for $E_{B-V}=1.0$ and $R_V=2.82$, shown as the black curve relative to the stellar continuum (horizontal line). The standard Cardelli et al.$\;$extinction law [19] is shown by the red curve for the same value of $R_V$, the gray area representing the depression of the stellar continuum by the effects of interstellar extinction for that law. The dashed line lying above the solid curve in the ultraviolet region represents forward scattered starlight (the dark gray area) by Rayleigh scattering according to Zagury [22]. Vertical lines denote the location of diffuse interstellar bands at 2175 \AA$\;$and 4430 \AA, as well as the {\it U}, {\it B}, and {\it V} bands for Johnson system photometry.}}
\label{fig1}
\end{figure*}

Trumpler demonstrated that extinction of starlight must exist, amounting to $\sim50$\% (0.7 magnitude in astronomical brightness) for every 1000 parsecs (= 1 kiloparsec, kpc) of distance in the Galactic plane [6], otherwise open cluster diameters would increase systematically with distance, a most unlikely situation. The extinction differs markedly from that value in some directions. Kapteyn, for example, earlier inferred 1.0 magnitude of extinction per kpc [7] from star counts in the direction of the Galactic centre, whereas NGC~1893 towards the anticentre contains stars displaying extnction of only $\sim1.2$ magnitude at a distance of $\sim4$ kpc [8]. Trumpler's study is of interest because it was the first instance in which photometry and spectroscopy of open clusters was used in conjunction with main-sequence fitting techniques to derive cluster reddenings and distances, in the process revealing information on the nature of yet another component of the Galaxy, interstellar dust. Subsequent refinements in the methods of studying open clusters resulted in modifications both to the variable-extinction technique and to cluster main-sequence fitting that helped establish mean properties for interstellar extinction as well as their spatial variations with location in the Galactic plane [8].

Studies of the wavelength ($\lambda$) dependence of interstellar extinction are tied to spectrophotometry of hot stars (those of spectral types O, B, and A), such a restriction being essential to reduce the alternate effect caused by masking of stellar continua by spectral lines, which increase in abundance and crowding for cooler stars of spectral types F, G, and K. In addition, extinction by Earth's atmosphere contaminates in systematic fashion photometric observations of stars made from ground-based facilities. Its effects must also be removed to isolate the effects of interstellar extinction (see [9]). It is now recognized from spectrophotometric studies of reddened stars [10,11,12,13,14,15] years ago that, at visible and near-infrared wavelengths, interstellar extinction depresses stellar continua in magnitude units in nearly linear fashion as a function of $\lambda^{-1}$, as seen in Fig.~14 of Nandy's study [15], as well as in the studies by Zagury [16], Zagury \& Turner [17], and in Fig.~1 below.

It is convenient to graph the effects of interstellar extinction on stellar continua using a colour difference diagram, which plots continuum depression in stellar magnitudes as a function of inverse wavelength ($\lambda^{-1}$, e.g. [18]), as illustrated in Fig.~\ref{fig1}. Here we follow the standard astronomical practice of plotting stellar spectra with wavelength increasing along the X-axis. It can be noted that the extinction is nearly linear as a function of $\lambda^{-1}$ from 0.8 $\mu$m$^{-1}$ ($\lambda = 1.2\;\mu$m) to 2.3 $\mu$m$^{-1}$ ($\lambda = 0.435\;\mu$m). Deviations from a linear law occur at shorter wavelengths, namely those lying below ``Nandy's knee'' [19] at $\lambda=0.435\;\mu$m (= 4350 \AA), where the ultraviolet interstellar extinction normally displays some curvature (Fig.~14 in [15], and Fig.~1). At such wavelengths the amount of extinction appears to be less than that expected from a simple continuation of the linear law at visible and infrared wavelengths. We expect no extinction at infinitely long wavelengths.

In the {\it UBV} photometric system of Johnson \& Morgan [20] the {\it B} (4365 \AA) and {\it V} (5475 \AA) bands lie on the linear part of the visible extinction relation, while the {\it U} (3500 \AA) band lies below Nandy's knee. The degree of linearity of the extinction curve can therefore be assessed, in the Johnson system, using the parameter $E_{U-B}/E_{B-V}$ describing the ratio of the reddening $E_{U-B}$, the difference in extinction in magnitudes between the {\it U} and {\it B} bands, to the similar amount of reddening $E_{B-V}$ describing the difference in extinction between the {\it B} and {\it V} bands. The ratio can also be written as $E_{U-B}/E_{B-V}=X+Y\times E_{B-V}$, the curvature term {\it Y} arising from the changing wavelength sensitivity of broad band filters on stellar continua as reddening increases. The parameter {\it X}, the slope of the reddening line [21], varies with position along the Galactic plane [8,22,23], while the curvature term {\it Y} is small. A study by Turner [24] established a mean value of $Y=0.021\pm0.006$ for local regions of the Galaxy from high quality photometric and spectroscopic data for O and B-type stars in restricted Galactic fields. Typical reddening slopes {\it X} appear variable, however, ranging from perhaps 0.55 to 0.84 along different directions in the Galactic plane. As will be demonstrated here, a linear extinction law of the type described by Nandy [15] produces $X=0.83$.

The nature of interstellar dust has been established from the wavelength dependence of interstellar extinction and polarization (linear and circular) in conjunction with the observed emission properties of warm dust clouds and simulations using Mie models for light scattering off spheroidal particles. In the unified model proposed by Li \& Greenberg [25], the extinction and reddening in the visible and infrared arises from collections of small amorphous silicate particles with overall dimensions similar to the wavelengths of optical light; an apparent dip in the ultraviolet extinction at 2175 \AA$\;$is attributed to molecular absorption produced by small carbonaceous grains (see the dip in the red curve in Fig.~\ref{fig1}), while a flattening of the apparent extinction in the far ultraviolet (short wavelength portion of the red curve) is attributed to very small graphite or silicate particles, or possibly polycyclic aromatic hydrocarbons (PAHs). In addition there are a number of diffuse interstellar bands (DIBs), of as-yet unspecified origin, superposed on stellar spectra [26], although it is unclear if the 2175 \AA$\,$bump should be included in the sample. One of the first DIBs to be discovered occurs at $4430$ \AA, with a maximum depth reaching $\sim15$\% of the continuum level in stars. Most others are found at longer wavelengths.

\begin{figure}[!t]
\center
\includegraphics[width=7cm]{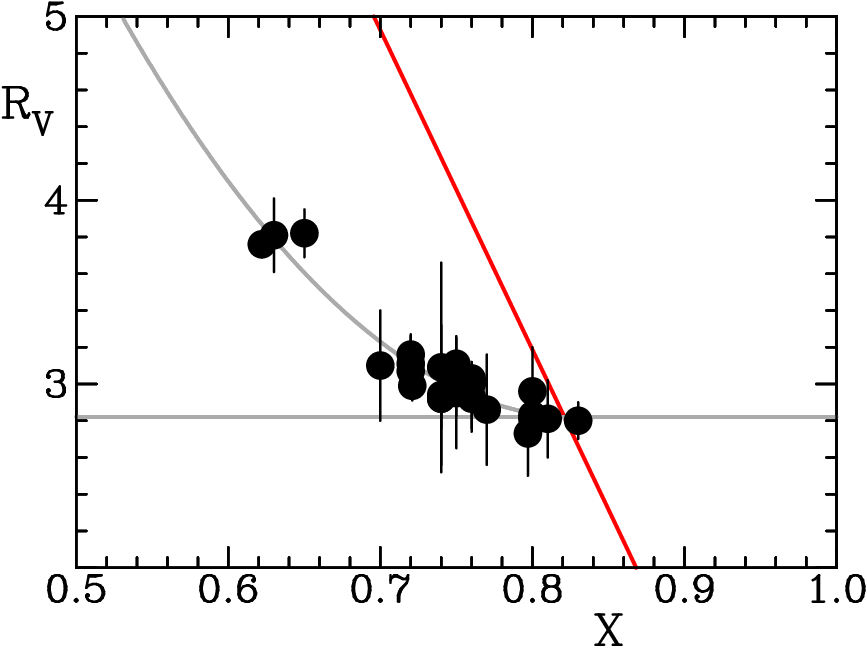}
\caption{\small{The variation of $R_V({\rm observed})$ with reddening slope {\it X} according to the data of Table~\ref{tab1}. The illustrated quadratic fit reaches minimum at $R_V=2.82$ for $X=0.83$. Shown for comparison is the linear trend (red line) expected from the formulae of Cardelli et al.$\;$[29].}}
\label{fig2}
\end{figure}

Correcting for the effects of interstellar reddening and extinction is essential for nearly all studies in Galactic astronomy tied to observations at optical wavelengths. Extinction corrections, e.g., $A_V$ for visual extinction, are made from a star's measured reddening, $E_{B-V}$, using the ratio of total-to-selective extinction, $R_V=A_V/E_{B-V}$ applicable to dust extinction towards the object. Observed values of $R_V$ are found to vary with Galactic location [8,24,27,28], as also noted by Cardelli et al.$\;$[29] when they derived empirical relations that appear to match interstellar extinction in the ultraviolet through infrared for different directions in the Galaxy. The Cardelli et al.$\;$extinction law [29] predicts, among other features, that variations in {\it X} and $R_V$ should be linearly related (red line in Fig.~\ref{fig2}). Perhaps because of ease of use, many researchers prefer to adopt mean reddening laws in their studies, typically for a value of $R_V=3.1$ (e.g., [30,31]).

The Cardelli et al.$\;$study [29] established that interstellar extinction, from the infrared to the far-ultraviolet, can be modeled using empirical polynomial functions with a dependence on only one parameter, $R_V$, in addition to reddening $E_{B-V}$. On the other hand, their extinction formulae are tied to complex polynomial fits (with up to seventh order terms) that have no physical meaning. In some cases the cited fits to observations are not particularly accurate (e.g., reddened stars with no 2175 \AA$\;$bump) and may be considerably improved if a dependence of the 2175 \AA$\;$bump on $E_{B-V}$ is introduced [32]. In addition, the decision by Cardelli et al.$\;$that $R_V$ is the free parameter of interstellar extinction does not appear to be judicious; it can be shown that a mathematical consequence of the Cardelli et al.$\;$[29] law is that $R_V$ must be constant [16].

An alternate view of the extinction law has been developed by Zagury [32], who suggests that the existence of linear extinction over the entire infrared through optical region and the one-parameter dependence of the Cardelli et al. [29] normalized extinction curves in the ultraviolet and far infrared are incompatible with potential variations in three independent types of particles from one interstellar dust cloud to another [25]. Zagury assumes that the lack of extinction relative to a simple $\lambda^{-1}$ relation observed below Nandy's knee does not result from a change in the properties of interstellar grains, but is the consequence of forward scattered starlight by interstellar gas clouds along the line of sight to the observer enhancing the observed amount of radiation from a star. As yet it is the only alternative to classical interstellar grain models. The very large amount of scattered starlight that is observed in the ultraviolet in the Zagury formulation [32], along with its  $\lambda^{-4}$ Rayleigh scattering intensity dependence, require that the scattered light be coherent scattering by hydrogen in the forward direction. An introduction to coherent scattering by hydrogen clouds in space is provided by Zagury \& Pellat-Finet [33], although there is as yet no satisfying theory on the subject.

In such a situation true interstellar extinction becomes a simple case of Mie scattering over the entire infrared to far ultraviolet wavelength region, with an extinction law described by $A(\lambda) \propto \lambda^{-p}$, as specified by the size distribution of the dust grains responsible. As an aside, atmospheric aerosols produce an identical dependence for their extinction [16,17]. A value of $p=1$, the simplest situation, would imply an invariant value of $R_V=4$ throughout the Galaxy [16,17], whereas observed values of $R_V\simeq3$ are typical of most Galactic fields [8]. Associated values of {\it p} for interstellar extinction must therefore lie in the range $1.3-1.4$ [17], which is still similar to what is observed for atmospheric aerosols. By implication the size distributions of atmospheric and interstellar dust particles must be governed by similar physical mechanisms. Anticipating the results presented later in this study, we argue that the true dependence for diffuse Galactic interstellar extinction is described by $A(\lambda) \propto \lambda^{-1.375}$, corresponding to $R_V=2.82$,  as illustrated in  Fig.~\ref{fig1}.

\begin{table}[!t]
\caption{Reddening and Extinction Data for Galactic Fields.}
\center
\begin{tabular}{lcccc}
\hline \hline
Region & {\it X} & $R_V$ & $\pm$s.e. & Source \\
\hline
Berkeley~58 &0.750 &2.95 &$\pm0.30$ &[44] \\
Berkeley~59 &0.800 &2.81 &$\pm0.09$ &[45] \\
NGC~129 &0.760 &3.01 &$\pm0.07$ &[8] \\
NGC~654 &0.760 &2.92 &$\pm0.16$ &[8] \\
IC~1805 &0.760 &3.01 &$\pm0.06$ &[8] \\
Alessi~95 &0.830 &2.80 &$\pm0.10$ &[46] \\
Pleiades &0.750 &3.11 &$\pm0.15$ &[8] \\
NGC~1647 &0.770 &2.86 &$\pm0.30$ &[8] \\
CV~Mon &0.760 &3.03 &$\pm0.07$ &[47] \\
NGC~2439 &0.760 &3.00 &$\pm0.10$ &[48] \\
Westerlund~2 &0.630 &3.81 &$\pm0.20$ &[49] \\
Ruprecht~91 &0.650 &3.82 &$\pm0.13$ &[50] \\
Lyng\aa~2 &0.720 &3.16 &$\pm0.10$ &[8] \\
NGC~5617 &0.720 &3.07 &$\pm0.11$ &[8] \\
Pismis~20 &0.740 &2.92 &$\pm0.16$ &[51] \\
NGC~6193 &0.760 &2.93 &$\pm0.19$ &[8] \\
Messier~4 &0.622 &3.76 &$\pm0.07$ &[43] \\
NGC~6611 &0.721 &2.99 &$\pm0.08$ &[27] \\
Trumpler~35 &0.740 &2.92 &$\pm0.10$ &[8] \\
Collinder~394 &0.700 &3.10 &$\pm0.10$ &[52] \\
NGC~6830 &0.720 &3.11 &$\pm0.16$ &[8] \\
NGC~6834 &0.750 &2.96 &$\pm0.13$ &[8] \\
SU~Cyg &0.740 &2.94 &$\pm0.38$ &[53] \\
P~Cyg &0.797 &2.73 &$\pm0.23$ &[54] \\
NGC~7062 &0.810 &2.81 &$\pm0.21$ &[8] \\
Trumpler~37 &0.800 &2.83 &$\pm0.12$ &[27] \\
NGC~7128 &0.800 &2.96 &$\pm0.24$ &[8] \\
NGC~7654 &0.750 &3.03 &$\pm0.15$ &[8]\\
NGC~7790 &0.740 &3.09 &$\pm0.57$ &[8] \\
\hline
\end{tabular}
\label{tab1}
\end{table}

An important aspect of the Zagury description of interstellar extinction [32] is how it affects the interpretation of observations in the ultraviolet. For the Cardelli et al. empirical formulation [29], as in the unified model of Li \& Greenberg [25], all extinction originates from interstellar dust particles in accordance with their size distribution, in conjunction with the spectroscopic properties of the constituents, which contribute specific absorption dips via atomic and molecular transitions, such as that observed at 2175 \AA,~the strongest of the diffuse interstellar bands (see Fig.~\ref{fig1}). According to the Zagury proposal [32], the extinction rise in the ultraviolet that generates the pseudo-continuum on which the 2175 \AA$\;$feature is superposed is an artifact of a separate process, namely forward scattered starlight that varies as $\lambda^{-4}$ [34,35,36,37,38], depicted schematically by the dashed line in Fig.~1. In this alternate scheme, interstellar extinction alone closely follows the $\lambda^{-1.375}$ relation depicted in Fig.~\ref{fig1}.

Forward scattering of starlight can be a relatively efficient process for particles with small dimensions relative to the wavelengths of light being scattered [33,35,36]. The origin of forward scattered starlight in the far ultraviolet is proposed by Zagury [16,32] to arise when light from a star passes through a distant cloud of hydrogen. Depending upon the relative distances of the star, cloud, and observer, a relatively thin cloud of hydrogen sufficiently distant from both a reddened star and Earth will redirect starlight along the line of sight to the star [33]. The proportion of forward scattered light can be more than an order of magnitude greater than the incident starlight in the forward direction, thereby accounting for an artificial $\lambda^{-4}$ continuum in the far ultraviolet.

In the Zagury interpretation of interstellar extinction curves [32] the 2175 \AA$\;$bump appears naturally as an interruption of the scattered starlight (note that the bottom of the 2175 \AA$\;$feature never dips below the $\lambda^{-1.375}$ extinction curve), but its physical origin is unclear. Since there is no minimum in the Rayleigh scattering cross-section for hydrogen, the bump could correspond to a resonance phenomenon associated with coherent scattering. Possibly it is related to the strong diffuse interstellar band (DIB) at 4430 \AA, about twice the bump's wavelength, and perhaps to the weaker DIB at 6614 \AA, roughly three times its wavelength. A recent suggestion by Sorokin \& Glownia [39] that the 2175 \AA$\;$feature can be created by two-photon absorption from molecular hydrogen (H$_2$) under conditions of induced resonance Rayleigh scattering may be relevant. It is no coincidence that H$_2$ has also been suspected for many years to explain the origin of the 4430 \AA$\;$feature (e.g., [40]). The strength of the DIB at 4430 \AA$\;$is known to correlate quite closely with the strength of the 2175 \AA$\;$bump [41,42], so a similar origin of both features is expected. A correlation between the relative strength of the DIB at 4430 \AA$\;$and {\it X} was also found by Wampler [21] fifty years ago, but was not investigated further.

The new interpretation of interstellar extinction explains a number of anomalies left unexplained by interstellar grain models and the Cardelli et al.$\;$formulation [29], in particular the existence of clearly reddened ($E_{B-V}>0$) stars that display no far ultraviolet rise or 2175 \AA$\;$absorption. Such stars display ultraviolet extinction that varies roughly as $\lambda^{-1.375}$, so the lack of a far ultraviolet rise or 2175 \AA$\;$absorption must be attributed to the lack of a hydrogen cloud sufficiently distant from the reddened star and Earth to generate forward-scattered starlight. Many such stars lie in directions away from the Galactic plane, e.g., towards the Magellanic Clouds [16,38], where only Galactic cirrus is capable of generating forward scattered starlight.
 
The new understanding also leads to interesting predictions regarding the optical reddening law and observed values of $R_V$. If scattered starlight is strong enough in the ultraviolet, it will gradually become more important in the optical wavelength region, first in the standard ultraviolet {\it U} band, and then, to a lesser extent, in the blue {\it B} band. It might even contribute a very minor component in the visual {\it V} band. The effect is to decrease the colour excesses $E_{U-B}$ and $E_{B-V}$ for a star systematically from their true values, more so for $E_{U-B}$, with a related effect on the observed reddening slope {{\it X}}. In similar fashion the value for the ratio of total-to-selective extinction $R_V({\rm observed})$ increases above its true value of 2.82 for interstellar extinction. Observed values of {\it X} should therefore decrease under such conditions, while observed values of $R_V$ will increase. The minimum of observed $R_V$ values should therefore correspond to the true value for interstellar extinction without a scattered light component, producing the best {\it p}-value that fits the simple interstellar grain extinction law of Zagury [32].

In this framework the scattered light component of extinction curves has a dependence on the square of the reddening ($E_{B-V}^2$). The relationship between {\it X} and $R_V$ should therefore be more complex than that expected from the Cardelli et al.$\;$formulation [29], namely quadratic rather than linear. An alternate version of such a relationship was also speculated upon by the lead author [27,28]. An increase in the scattered light component for a star is also predicted to result in stronger 2175 \AA$\;$and 4430 \AA$\;$absorption bands, which must increase in strength in proportion to a decrease in the observed extinction slope {{\it X}}. Such a dependence for the 4430 \AA$\;$feature was indeed noted by Wampler fifty years ago [21], as noted earlier, but never investigated further. The present investigation was therefore designed as a follow-up to Wampler's original study, specifically aimed at tying together the various pieces of observational evidence to test the ideas proposed by Zagury [32] regarding a new description of interstellar extinction.

\section{{\rm \footnotesize DEPENDENCE OF $R_V({\rm observed})$ ON {\it X}}}
Establishing an estimate of reddening slope {\it X} for any star field is a challenging task that requires high quality photometry and spectral types for stars in the field. By comparison, the Cardelli et al.$\;$relationship [29] in the optical and infrared is tied mainly to colour difference data for stars, which can introduce systematic effects in some situations [27]. The use of high quality data for stars in restricted star fields is just as effective [24], and helped to generate the observational correlation between $R_V$ (observed) and {\it X} noted by Turner [28]. At the time of that discovery it was intended to confirm the correlation using results acquired in an earlier study [8], but the variable-extinction data in that study were of mixed quality, limiting possible conclusions.

For the present study we re-examined the variable-extinction results from the Turner study [8] to isolate those of optimum quality, and added results from subsequent studies by the lead author, typically of open clusters containing Cepheid variables. We also included results for the globular cluster M14 by Hendricks et al. [43], without their adjustment to the Cardelli et al.$\;$scheme, since M14 is one of only a few clusters in the well-studied field of Upper Scorpius affected by differential reddening [24,27]. A compilation of the results is given in Table~\ref{tab1}, and the data are plotted in Fig.~\ref{fig2} as a dependence of $R_V$ (observed) on {\it X}.

The observational data of Fig.~\ref{fig2} appear to follow a quadratic trend rather than the linear relation derived by Turner [28] using far fewer data points. They also deviate from the trend expected from the Cardelli et al.$\;$reddening law [29], shown as the red line in Fig.~\ref{fig2}. A least squares quadratic solution to the data has a minimum at $R_V ({\rm observed})=2.82\pm0.06$ and $X=0.83\pm0.01$, the result adopted in Fig.~\ref{fig1}. By implication the true value of $R_V$ for interstellar extinction is 2.82, and is associated with an interstellar reddening law of slope $X=0.83$. Smaller reddening slopes {\it X} and larger values of $R_V$(observed) must be artifacts of enhanced forward scattered starlight in the ultraviolet permeating disproportionately into the optical wavelength bands.

The data of Fig.~\ref{fig2} also argue that minimum values of $R_V$ (observed) in the Galaxy must be around 2.8 with maximum values of perhaps 4. The latter can be accommodated by modifying the exponent of the $\lambda^{-1.375}$ extinction law to $\lambda^{-1}$ ([16], see also [49]), but that is merely a cosmetic correction to adjust for the scattered starlight component superposed on the optical continuum. The true interstellar extinction relation $A(\lambda)\propto\lambda^{-1.375}$ is presumably unchanged. In addition, the observed range of reddening slope {\it X} from 0.62 to 0.83 may represent the maximum variation possible for apparent extinction in the Milky Way by interstellar dust clouds intermingled with clouds of hydrogen, the true reddening slope remaining unchanged at 0.83. By implication the large values of $R_V({\rm observed})\simeq5$ cited for stars in some H~II regions must represent unique circumstances associated with neutral extinction that enhances the overall extinction towards an object with minimum influence on the reddening slope {\it X}.

Reliance on the Cardelli et al.$\;$relations [29] for correcting for the effects of interstellar extinction in the Galaxy can no longer be supported according to the results of Fig.~\ref{fig2}. The Cardelli et al.$\;$law [29] predicts variations in both reddening slope {\it X} and ratio of total-to-selective absorption $R_V({\rm observed})$ that exceed the range of observed values (Table~\ref{tab1}) and deviate markedly from them (Fig.~\ref{fig2}). Given the other weaknesses pointed out in $\S1$, corrections for interstellar extinction in Galactic studies in future must use relationships more suitable to the specific regions under investigation.

\section{{\rm \footnotesize THE $\lambda4430$ DEPENDENCE}}
In order to replicate the results of Wampler [21], we collected equivalent width data on 4430 \AA$\;$band absorption for stars in several open cluster fields for which the reddening slope {\it X} is reliably known. For the clusters Alessi~95 and Berkeley~87 the data were measured from intensity-normalized CCD spectra obtained for studies of the clusters [46,55] . For Westerlund~2 the data were obtained from similarly-normalized spectra published by Rauw et al.$\;$[56]. Additional data were taken from the digital atlas of OB star spectra published by Walborn \& Fitzpatrick [57], and from scanner observations of the 4430 \AA$\;$feature by Underhill [58], Stoeckly \& Dressler [59], and Wampler [60], with the $A_c$ measures (in percentages) catalogued there converted to the equivalent width of the 4430 \AA$\;$feature, $W(\lambda4430)$, using the formula $W=0.2\;A_c$ given by Underhill [58]. Both data types (equivalent widths and $A_c$ measures) were available for stars in NGC~2244, and gave essentially identical results, indicating that the photographic spectra of the 1950s were reasonably accurately reduced to an intensity scale by the techniques in use at the time.

\begin{figure}[!t]
\center
\includegraphics[width=7cm]{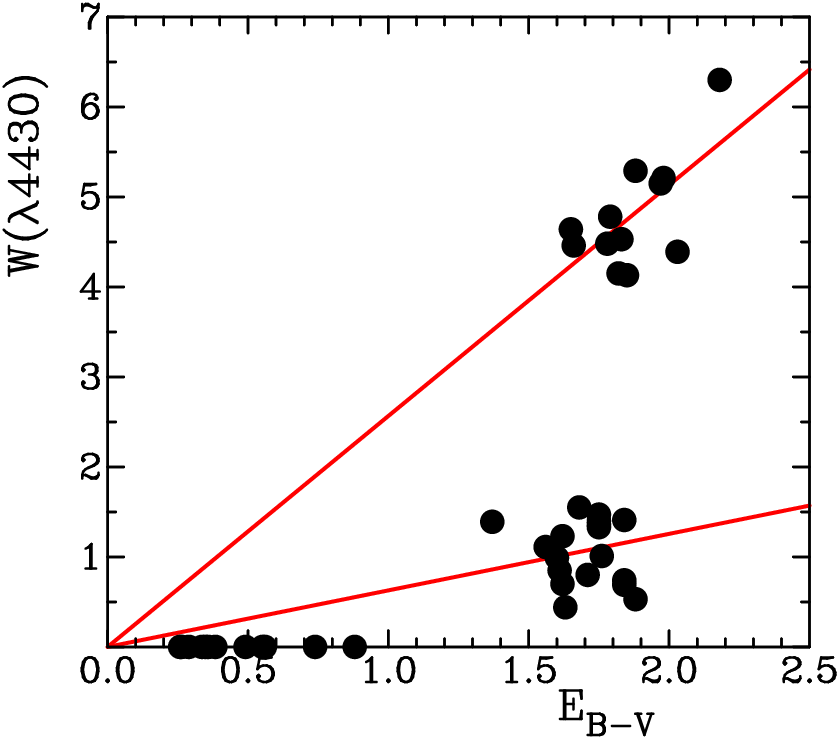}
\includegraphics[width=7cm]{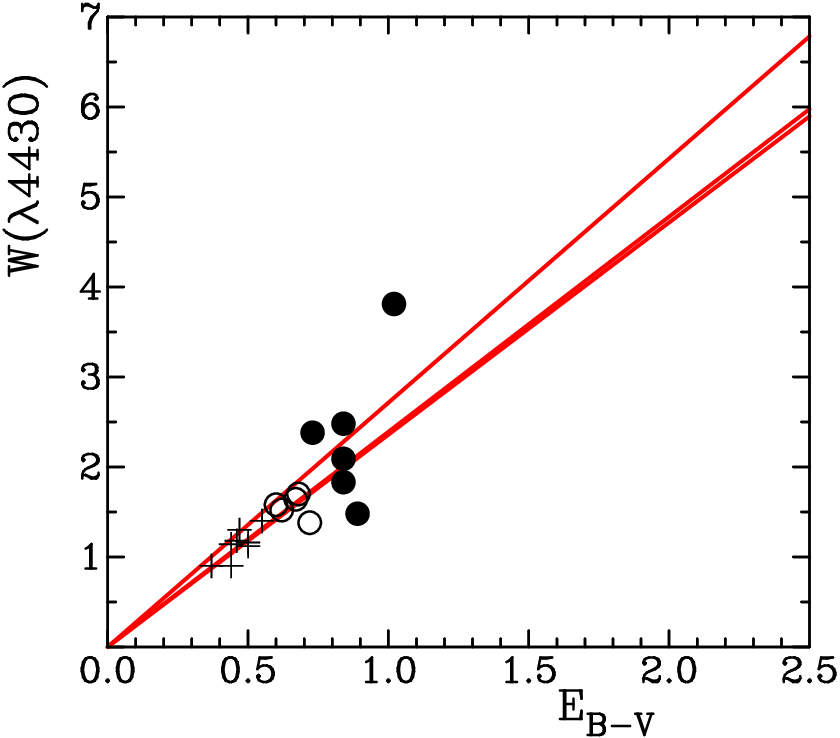}
\caption{\small{Measured equivalent widths for $\lambda4430$ as a function of colour excess $E_{B-V}$ for stars in (top portion) Westerlund~2 (upper), Berkeley~87 (middle), and Alessi~95 (bottom), and (lower portion) IC~1805 (filled circles), IC~4996 (open circles), and NGC~2244 (plus signs). Red lines denote mean relations for each cluster.}}
\label{fig3}
\end{figure}

The data for individual stars in six open cluster fields are shown in Fig.~\ref{fig3}, while mean values of equivalent width ratio $W(\lambda4430)/E_{B-V}$ and {\it X} for the fields analyzed are shown as sloped red lines in the figure and are summarized in Table~\ref{tab2}. There is no evidence for $\lambda4430$ absorption in stars belonging to Alessi~95; many of the stars are A-type objects in which normal spectral lines of Fe~I and Fe~II are prominent in that wavelength interval.

Correlations between $W(\lambda4430)$ and $E_{B-V}$ have long been recognized [21,26,58,59,60]. But the scatter displayed in general plots of the parameters is typically large, masking the exact functional dependence. According to Fig.~\ref{fig3} such scatter must be the result of different dependences throughout the Galaxy, according to the reddening slope {\it X} towards the direction studied.

\begin{figure}[h]
\center
\includegraphics[width=7cm]{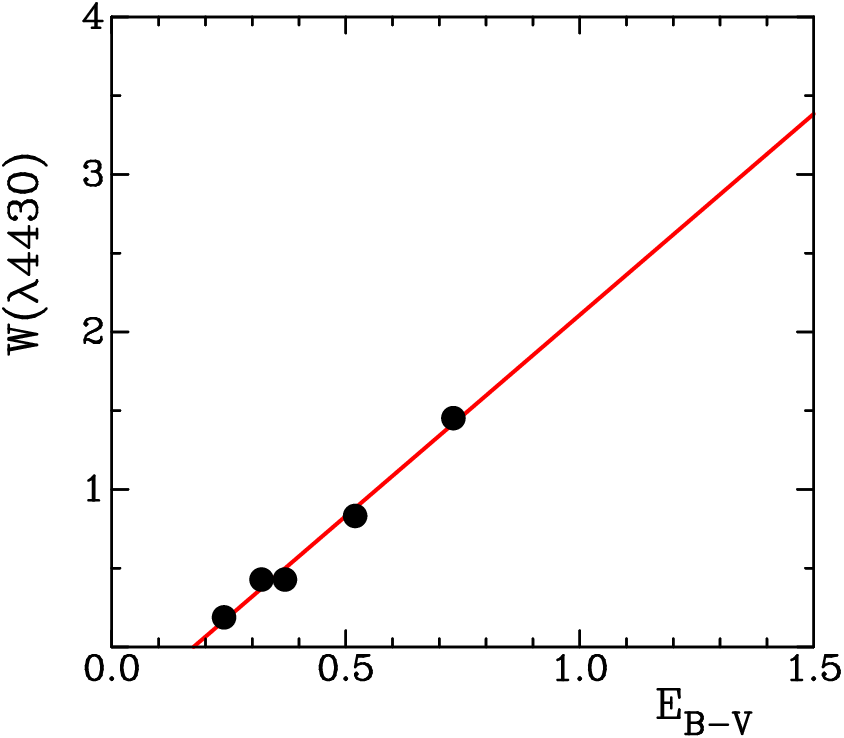}
\caption{\small{Measured equivalent widths for $\lambda4430$ as a function of colour excess $E_{B-V}$ for stars in Tr~16/Cr~228. Note the zero-point offset to $E_{B-V}=0.175$.}}
\label{fig4}
\end{figure}

The Carina field of Trumpler~16/Collinder~228 presents an interesting puzzle, as indicated by the data of Fig.~\ref{fig4}. A trend of increasing $W(\lambda4430)$ with $E_{B-V}$ is indicated, but is offset from the normal zero-point of $W(\lambda4430)=0$ at $E_{B-V}=0$ by a small amount in colour excess; the trend begins at a reddening of $E_{B-V}=0.175$. It apparently indicates that an initial reddening of $E_{B-V}=0.175$ for stars in Tr~16/Cr~228 is described by extinction without a scattered light component, much like AzV$\:$18 in the Small Magellanic Cloud [22,28], which incidentally is reddened by the same amount. It suggests that the initial reddening for cluster stars originates from interstellar matter close to the stars [22]. Thereafter the stars are affected by a forward-scattered component of starlight that must originate in a more distant hydrogen cloud lying between the Sun and the clusters.

\begin{figure}[!t]
\center
\includegraphics[width=7cm]{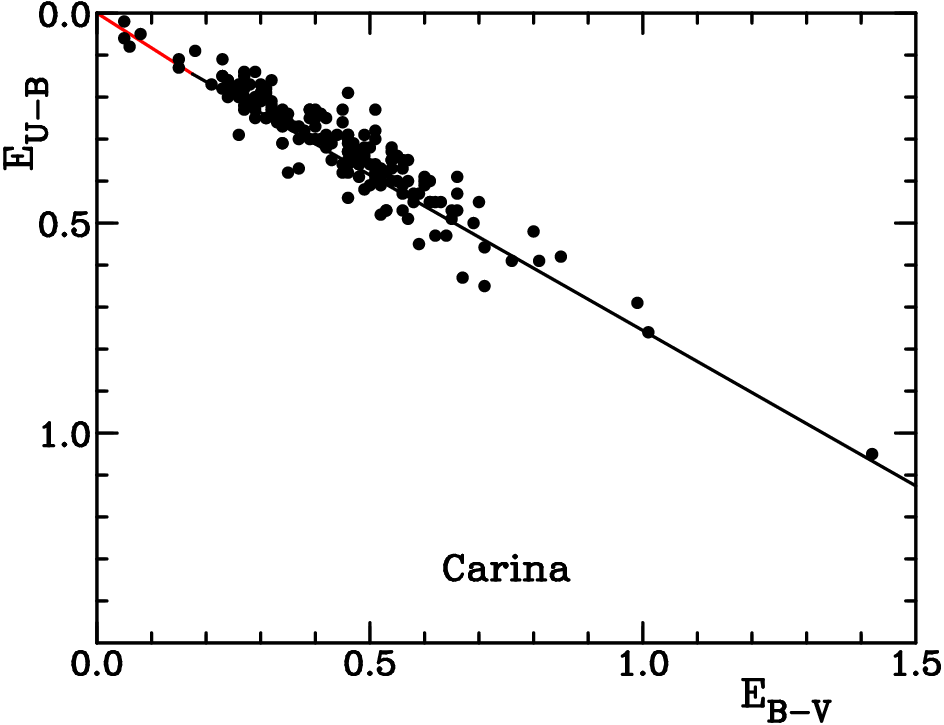}
\caption{\small{Measured colour excesses for stars in Tr~16/Cr~228 and surrounding fields. Initial reddening of $E_{B-V}=0.175$ is described by $X=0.83$ (red line), and larger reddening by $X=0.74$ (black line), in reasonably good agreement with most of the data. A reddening slope of $X=0.63$ applies in some adjacent fields, and may affect the data for some stars.}}
\label{fig5}
\end{figure}

Determining the effective reddening slope for the same stars is somewhat challenging. A study of the Carina Nebula clusters Trumpler~16 and Collinder~228 by Turner \& Moffat [61] derived a reddening slope of $X\simeq0.75$ for the field. But a study of the anomalous extinction towards Carina by Turner [62] noted the dificulties of assigning a specific value to the reddening slope {\it X} of stars in the region. If the initial reddening of $E_{B-V}=0.175$ applies to cluster stars, then we expect that $X=0.83$ is applicable for that initial reddening. Thereafter the reddening slope decreases as the scattered light component dominates, and will average $X\simeq0.75$ only if $X=0.74$ for stars of larger reddening. A plot of colour excesses for stars in the field of the two clusters is presented in Fig.~\ref{fig5}, which replicates the data in Fig.~8 of Turner [62] and demonstrates that the interpretation presented here is reasonably consistent with the observational data.

\begin{table}[!t]
\caption{Equivalent Width Data for $\lambda4430$ in Open Clusters.}
\center
\begin{tabular}{lccc}
\hline \hline
Region & {\it X} & $W(\lambda4430)/E_{B-V}$ & $\pm$s.d. \\
\hline
Westerlund~2 &0.62 &2.566 &$\pm0.238$ \\
Sco-Oph &0.63 &2.187 &$\pm0.512$ \\
IC~1590 &0.73 &2.286 &$\cdots$ \\
IC~4996 &0.74 &2.390 &$\pm0.138$ \\
Tr~16/Cr~228 &0.74 &2.553 &$\pm0.163$ \\
NGC~7380 &0.75 &2.768 &$\pm0.165$ \\
IC~1805 &0.75 &2.713 &$\pm0.754$ \\
NGC~2244 &0.77 &2.254 &$\pm0.078$ \\
Cyg~OB2 &0.80 &1.406 &$\pm0.225$ \\
Trumpler~37 &0.80 &1.321 &$\cdots$ \\
Polaris &0.80 &1.079 &$\cdots$ \\
Berkeley~87 &0.81 &0.628 &$\pm0.221$ \\
Alessi~95 &0.83 &0.000 &$\cdots$ \\
\hline
\end{tabular}
\label{tab2}
\end{table}

\begin{figure}[!t]
\center
\includegraphics[width=7cm]{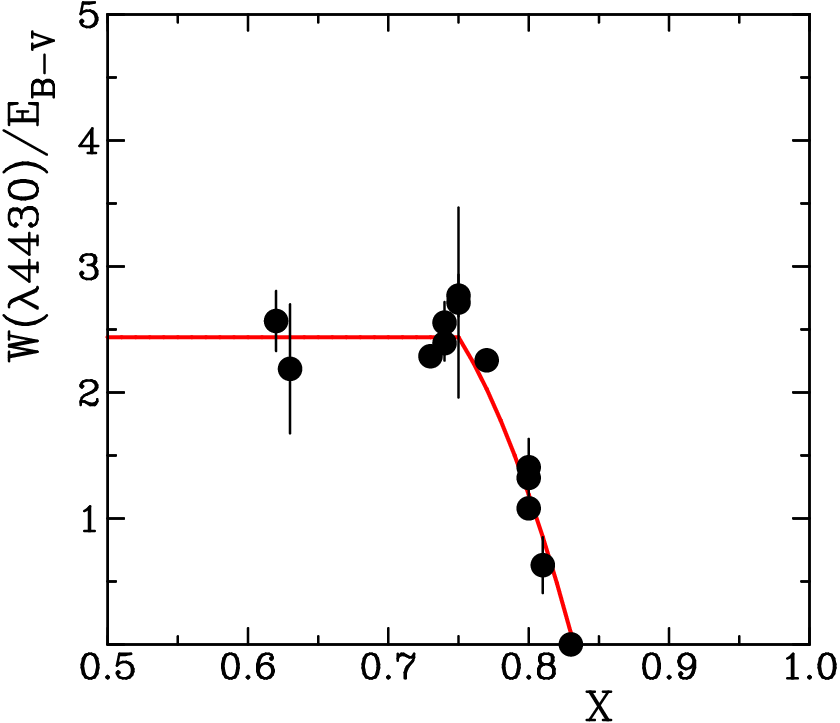}
\caption{\small{Observed dependence of normalized $\lambda4430$ equivalent widths as a function of reddening slope {\it X}. The gray quadratic for $X\ge0.75$ is a best fit to the data, with no $\lambda4430$ absorption at $X=0.83$ and saturation adopted for $X\le0.75$.}}
\label{fig6}
\end{figure}

The general variation of normalized $\lambda4430$ absorption summarized in Table~\ref{tab2} as equivalent width ratio $W(\lambda4430)/E_{B-V}$ is plotted with respect to our adopted independent variable, colour excess slope {\it X}, in Fig.~\ref{fig6}. The fact that stars for which $X=0.83$ display no evidence for $\lambda4430$ absorption is expected from the discussion by Zagury [22] about the strength of the 2175 \AA$\;$feature, which correlates directly with 4430 \AA$\;$absorption according to Nandy \& Thompson [29] and Wu et al.$\;$[30]. That is substantiated by International Ultraviolet Explorer (IUE) spectra for stars in the fields studied. Most stars near SU~Cas (only a few were observed by IUE) exhibit either weak or no 2175 \AA$\;$feature, consistent with the $\lambda4430$ equivalent widths, while the few stars observed by IUE in the Carina cluster fields exhibit strong 2175 \AA$\;$absorption, as expected.

The initial increase in normalized $\lambda4430$ absorption with decreasing reddening slope {\it X} in Fig.~\ref{fig6} is expected from the discussion associated with Fig.~\ref{fig1}. As the forward-scattered starlight component strengthens in the near-ultraviolet {\it U} band, it encroaches upon the reddened stellar continuum in the optical region (the 4430 \AA$\;$DIB falls in the {\it B} band). Around $X=0.75$ the normalized $\lambda4430$ absorption appears to saturate, and remains at the same level as {\it X} decreases further. That must relate directly to the mechanism that creates the 2175 \AA$\;$and 4430 \AA$\;$features. Note that the observed colour excesses $E_{B-V}$ for $X\le0.75$ are systematically underestimated from their true values in this scheme, so the saturation depicted in Fig.~\ref{fig6} is, if anything, underestimated. It would be interesting to establish if there is a corresponding saturation in the strength of the 2175 \AA$\;$feature for stars lying in regions of small reddening slope {\it X}. Otherwise, the linear trend seen at larger reddening slopes is consistent with the results found by Wampler [22,23], despite small differemces in the adopted values of {\it X} for each field.  

It may be possible to include additional data in Fig.~\ref{fig6} using measures of $W(\lambda4430)$ for stars lying in other regions where the reddening slope is well established. The main hindrance to such research is finding suitable spectroscopic data or measures of $\lambda4430$ equivalent width, in conjunction with high quality photometry and spectral types for the establishment of reddening slope {\it X}. The implied correlation between normalized $\lambda4430$ absorption and relative strength of the 2175 \AA$\;$ feature proposed here is worth investigating further, but is difficult to confirm with the ultraviolet observations available. The concept does remain a distinct possibility, however.

\section{{\rm \footnotesize DISCUSSION}}
The nature of interstellar extinction in the Galaxy is tested here using empirical observational data related to variations in reddening slope {\it X} ($=E_{U-B}/E_{B-V}$ for small reddening), observed ratio of total-to-selective extinction $R_V$, and equivalent width $W(\lambda4430)$ for stars in well-studied Galactic fields, typically regions of open clusters where the properties of the extinction do not vary appreciably from one star to another. The derived quadratic variation of $R_V({\rm observed})$ with {\it X} is presumably related to increasing amounts of forward-scattered starlight in the ultraviolet affecting the optical colour excesses $E_{U-B}$ and $E_{B-V}$, as well as $R_V({\rm observed})$, in systematic fashion, the effect increasing in direct proportion to the amount of forward-scattered starlight in the ultraviolet. As noted by Zagury [32], the colour excesses for stars displaying such effects are undoubtedly underestimated from what they would be in the absence of scattered starlight in the ultraviolet. The observed dependence differs markedly from expectations predicted by the Cardelli et al.$\;$relationship for interstellar extinction [29].

An associated effect on equivalent widths $W(\lambda4430)$ for the interstellar 4430 \AA$\;$feature is expected because of the direct correlation between the strength of the 2175 \AA$\;$feature with $W(\lambda4430)$ [41,42]. The observed effect (Fig.~\ref{fig6}) is mostly as expected, with normalized $W(\lambda4430)$ absorption increasing directly with decreasing {\it X} from $W(\lambda4430)/E_{B-V}=0$ at $X=0.83$. But there is a subsequent saturation of the normalized absorption near $X=0.75$ that, while expected, is not fully explained by the physical concepts of extinction proposed by Zagury [32]. Presumably it is associated with the mechanism by which absorption at 2175 \AA$\;$originates. In the original Zagury proposition [32] a resonance mechanism for the forward scattering of starlight by clouds of hydrogen was proposed to explain the 2175 \AA$\;$absorption, but apparently such considerations are now focusing on a possible origin from interstellar H$_2$ molecules, as suggested by Sorokin \& Glownia [39]. Either mechanism may also apply to absorption at 4430 \AA, i.e. resonance in interstellar hydrogen or absorption by H$_2$ molecules, given the obvious correlation between the two.

A major repercussion of the present findings is that use of the Cardelli et al.$\;$[29] relations to correct for interstellar extinction is no longer justified. Yet the same prescription is typical of the corrections made for extinction towards Cepheids in other galaxies. Although total extinctions for dust obscuration are made from observations over the visual and infrared {\it VRI} bands in order to minimize potential variations from the Galactic mean, it should be noted that dust extinction described by a $A(\lambda)\propto \lambda^{-1.375}$ absorption model predicts different intrinsic infrared colours from those usually adopted, as well as less total extinction $A_V$. The differences might appear to be small, but tests indicate that they are significant [63]. For example, the 2175 \AA$\;$ feature observed towards galaxies at high Galactic latitude is generally weak of absent, despite non-zero reddenings for the galaxies [64,65], implying that a value of $R_V=2.82$ is applicable for them.

By implication, the derived distances to galaxies containing Cepheid variables are likely underestimated in studies to date. Since the same galaxies serve as calibrators for the extragalactic distance scale, the observational determination of the local Hubble constant $H_0$ from such studies may contain systematic effects. That will influence the accuracy of $H_0$ expected by the SH0ES collaboration [66], for example, and possibly explain why recent values such as $H_0=74.3\pm2.1$ km s$^{-1}$ Mpc$^{-1}$ obtained by the Carnegie Hubble team using Cepheids [67] are systematically larger than values of $H_0=67.3\pm1.2$ km s$^{-1}$ Mpc$^{-1}$ obtained by the Planck collaboration [68] or $H_0=68.9\pm7.1$ km s$^{-1}$ Mpc$^{-1}$ obtained from UGC~3789 by Reid et al. [69]. The proper treatment of interstellar and extragalactic extinction is an important consideration in such comparisons.
 
\section*{{\rm \footnotesize ACKNOWLEDGEMENTS}}
\scriptsize{We are grateful to the director and staff of the Dominion Astrophysical Observatory, Herzberg Institute of Astrophysics, National Research Council of Canada, for providing observing time and support for spectroscopic observations with the 1.85m Plaskett Telescope, and to Frederic Zagury for useful comments on original versions of the paper.}


\begin{thebibliography}{}\setlength{\itemsep}{-1.5mm}
\bibitem{1}[1] Barnard, E.~E., Popular~Astron. {\bf 14}, 579 (1906).
\bibitem{2}[2] Kapteyn, J.~C., Astrophys.~J. {\bf 29}, 46 (1909a).
\bibitem{3}[3] Kapteyn, J.~C., Astrophys.~J. {\bf 30}, 284 (1909b).
\bibitem{4}[4] Comstock, G.~C., Astrophys.~J. {\bf 31}, 270 (1910).
\bibitem{5}[5] Trumpler, R.~J., Lick~Obs.~Bull. {\bf 14}, 154 (1930a).
\bibitem{6}[6] Trumpler, R.~J., Publ.~Astron.~Soc.~Pacific {\bf 42}, 214 (1930b).
\bibitem{7}[7] Kapteyn, J.~C., Astrophys.~J. {\bf 40}, 187 (1914).
\bibitem{8}[8] Turner, D.~G., Astron.~J. {\bf 81}, 1125 (1976).
\bibitem{9}[9] Hardie, R.~H., in {\it Astronomical~Techniques}, Ed. W.~A. Hiltner, University of Chicago Press: Chicago, Chapt. {\bf 8}, p. 178 (1962).
\bibitem{10}[10] Hall, J.~S., Astrophys.~J. {\bf 85}, 145 (1937).
\bibitem{11}[11] Greenstein, J.~L., Harvard~Coll.~Obs.~Circ. {\bf 422}, 1 (1937).
\bibitem{12}[12] Greenstein, J.~L., Astrophys.~J. {\bf 87}, 151 (1938).
\bibitem{13}[13] Stebbins, J., \& Whitford, A.~E., Astrophys.~J. {\bf 98}, 20 (1943).
\bibitem{14}[14] Whitford, A.~E., Astrophys.~J. {\bf 107}, 102 (1948).
\bibitem{15}[15] Nandy, K., Publ.~Roy.~Obs.~Edinburgh {\bf 3}, 142 (1964).
\bibitem{16}[16] Zagury, F., Astron.~Nachr. {\bf 333}, 160 (2012).
\bibitem{17}[17] Zagury, F., \& Turner, D.~G., Astron.~Nachr. {\bf 333}, 640 (2012).
\bibitem{18}[18] Whitford, A.~E., Astron.~J. {\bf 63}, 201 (1958).
\bibitem{19}[19] Wickramasinghe, C., Observatory {\bf 118}, 398 (1998).
\bibitem{20}[20] Johnson, H.~L., \& Morgan, W.~W., Astrophys.~J. {\bf 117}, 313 (1953).
\bibitem{21}[21] Wampler, E.~J., Astrophys.~J. {\bf 137}, 1071 (1963).
\bibitem{22}[22] Wampler, E.~J., Astrophys.~J. {\bf 134}, 861 (1961).
\bibitem{23}[23] Wampler, E.~J., Astrophys.~J. {\bf 136}, 100 (1962).
\bibitem{24}[24] Turner, D.~G., Astron.~J. {\bf 98}, 2300 (1989).
\bibitem{25}[25] Li, A., \& Greenberg, J.~M., Astron.~Astrophys. {\bf 323}, 566 (1997).
\bibitem{26}[26] Herbig, G.~E., Ann.~Rev.~Astron.~Astrophys. {\bf 33}, 19 (1995).
\bibitem{27}[27] Turner, D.~G., Rev.~Mex.~Astron.~Astrophys. {\bf 29}, 163 (1994).
\bibitem{28}[28] Turner, D.~G., in {\it The Origins, Evolution, and Destinies of Binary Stars in Clusters}, Eds. E.~F. Milone \& J.C. Mermilliod, Astron.~Soc.~Pacific~Conf.~Series {\bf 90}, 443 (1996a).
\bibitem{29}[29] Cardelli, J.~A., Clayton, G.~C., \& Mathis, J.~S., Astrophys.~J. {\bf 345}, 245 (1989).
\bibitem{30}[30] FitzGerald, M.~P., Astron.~Astrophys. {\bf 4}, 234 (1970).
\bibitem{31}[31] Rieke, G.~H., \& Lebofsky, M.~J., Astrophys.~J. {\bf 288}, 618 (1985).
\bibitem{32}[32] Zagury, F., Astron.~Nachr. {\bf 334}, 1107 (2013).
\bibitem{33}[33] Zagury, F., \& Pellat-Finet, P., Optics~Commun. {\bf 285}, 4001 (2012).
\bibitem{34}[34] Zagury, F., New~Astron, {\bf 6}, 403 (2001a).
\bibitem{35}[35] Zagury, F., New~Astron, {\bf 6}, 415 (2001b).
\bibitem{36}[36] Zagury, F., New~Astron, {\bf 7}, 117 (2002).
\bibitem{37}[37] Zagury, F., New~Astron, {\bf 10}, 237 (2005).
\bibitem{38}[38] Zagury, F., Astrophys.~Space~Sci. {\bf 312}, 113 (2007).
\bibitem{39}[39] Sorokin, P.~P., \& Glownia, J.~H., arXivastro-ph/0608092 (2006).
\bibitem{40}[40] Zimmermann, H., Astrophys.~Space~Sci. {\bf 84}, 505 (1982).
\bibitem{41}[41] Nandy, K., \& Thompson, G.~I., Mon.~Not.~Roy.~Astron.~Soc. {\bf 172}, 237 (1975).
\bibitem{42}[42] Wu, C.-C., York, D.~G., \& Snow, T.~P., Astron.~J. {\bf 86}, 755 (1981).
\bibitem{43}[43] Hendricks, B., Stetson, P.~B., VandenBerg, D.~A., \& Dall'Ora, M., Astron.~J. {\bf 144}, 25 (2012).
\bibitem{44}[44] Turner, D.~G., Forbes, D., English, D., Leonard, P.~J.~T., Scrimger, J.~N., Wehlau, A.~W., Phelps, R.~L., Berdnikov, L.~N., \& Pastukhova, E.~N., Mon.~Not.~Roy.~Astron.~Soc. {\bf 388}, 444 (2008).
\bibitem{45}[45] Majaess D.~J., Turner D.~G., Lane D.~J., \& Moncrieff, K.~A., J.~American~Assoc.~Var.~Star~Obs. {\bf 36}, 90 (2008).
\bibitem{46}[46] Turner, D.~G., Majaess, D.~J., Lane, D.~J., Balam, D.~D., Gieren, W.~P., Storm, J., Forbes, D.~W., Havlen, R.~J., \& Alessi, B., Mon.~Not.~Roy.~Astron.~Soc. {\bf 422}, 2501 (2012).
\bibitem{47}[47] Turner, D.~G., Pedreros, M.~H., \& Walker, A.~R., Astron.~J., {\bf 115}, 1958 (1998).
\bibitem{48}[48] Turner, D.~G., Astron.~J. {\bf 82}, 805 (1977).
\bibitem{49}[49] Carraro, G., Turner, D.~G., Majaess, D.~J., \& Baume, G., Astron. Astrophys. {\bf 555}, A50 (2013).
\bibitem{50}[50] Turner, D.~G., Forbes, D., van den Bergh, S., Younger, P.~F., \& Berdnikov, L.~N., Astron.~J. {\bf 130}, 1194 (2005).
\bibitem{51}[51] Turner, D.~G., Astron.~J. {\bf 111}, 828 (1996b).
\bibitem{52}[52] Turner, D.~G., \& Pedreros, M., Astron.~J. {\bf 90}, 1231 (1985).
\bibitem{53}[53] Turner, D.~G., Mandushev, G.~I., \& Welch, G.~A., Astron.~J. {\bf 113}, 2104 (1997).
\bibitem{54}[54] Turner, D.~G., Welch, G., Graham, M, Fairweather, D., Horsford, A., Seymour, M., \& Feibelman, W., J.~American~Assoc.~Var.~Star~Obs. {\bf 29}, 73 (2001).
\bibitem{55}[55] Turner, D.~G., Majaess, D.~J., Lane, D.~J., \& Balam, D.~ D., Bull.~American~Astron.~Soc. {\bf 42}, 566 (2010).
\bibitem{56}[56] Rauw, G., Manfroid, J., Gosset, E., Naz\'{e}, Y., Sana, H., De Becker, M., Foellmi, C., \& Moffat, A.~F.~J., Astron.~Astrophys. {\bf 463}, 981 (2007).
\bibitem{57}[57] Walborn, N.~R., \& Fitzpatrick, E.~L., Publ.~Astron.~Soc.~Pacific {\bf 102}, 379 (1990).
\bibitem{58}[58] Underhill, A.~B., Publ.~Dom.~Astrophys.~Obs. {\bf 10}, 201 (1956).
\bibitem{59}[59] Stoeckly, R., \& Dressler, K., Astrophys.~J. {\bf 139}, 240 (1964).
\bibitem{60}[60] Wampler, E.~J., Astrophys.~J. {\bf 144}, 921 (1966).
\bibitem{61}[61] Turner, D.~G., \& Moffat, A.~F.~J., Mon.~Not.~Roy.~Astron.~Soc. {\bf 192}, 283 (1980).
\bibitem{62}[62] Turner, D.~G., Astron.~Nachr. {\bf 338}, 303 (2012).
\bibitem{63}[63] Turner, D.~G., 224th Meeting of the American Astronomical Society 318.09 (2014).
\bibitem{64}[64] Bianchi, L., Clayton, G.~C., Bohlin, R.~C., Hutchings, J. B., \& Massey, P., Astrophys.~J. {\bf 471}, 203 (1996).
\bibitem{65}[65] Pitman, K.~M., Clayton, G.~C., \& Gordon, K.~D.., Astrophys.~J. {\bf 112}, 537 (2000).
\bibitem{66}[66] Macri, L.~M., \& Riess, A.~G., and the SH0ES Team, AIP~Conf.~Proc. {\bf 1170}, 23 (2009).
\bibitem{67}[67] Freedman, W.~L., Madore, B.~F., Scowcroft, V., Burns, C., Monson, A., Persson, S.~E., Seibert, M., \& Rigby, J., Astrophys.~J. {\bf 758}, 24 (2012).
\bibitem{68}[68] Ade, P.~A.~R., et al., arXiv:1303.5076 (2014).
\bibitem{69}[69] Reid M.~J., Braatz, J.~A., Condon, J.~J., Lo, K.~Y., Kuo, C.~Y., Impellizzeri, C.~ M.~V., \& Henkel, C., Astrophys.~J. {\bf 767}, 154 (2013).
\end{thebibliography}
\end{document}